\documentclass[prb,twocolumn]{revtex4-1}

\usepackage{amsmath,amsfonts,amssymb}
\usepackage{graphicx,color}
\usepackage{hyperref}
\usepackage{epstopdf}
\begin{document}

\title{Implementation of pairwise longitudinal coupling in a three-qubit superconducting circuit}

\author{Tanay Roy$^{1}$, Suman Kundu$^{1}$, Madhavi Chand$^{1}$, Sumeru Hazra$^{1}$}
\author{N. Nehra$^{1}$}
\altaffiliation{Current address: Dept. of Physics, University of Wisconsin-Madison, Madison, WI 53706, USA}
\author{R. Cosmic$^{1}$}
\altaffiliation{Current address: Research Center for Advanced Science and Technology (RCAST), The University of Tokyo, Meguro-ku, Tokyo 153-8904, Japan}
\author{A. Ranadive$^{1}$}
\author{Meghan P. Patankar$^{1}$, Kedar Damle$^{2}$}
\author{R. Vijay$^{1}$}
\email{Corresponding author: r.vijay@tifr.res.in}

\affiliation{$^{1}$Department of Condensed Matter Physics and Materials Science, Tata Institute of Fundamental Research, Homi Bhabha Road, Mumbai 400005, India}
\affiliation{$^{2}$Department of Theoretical Physics, Tata Institute of Fundamental Research, Homi Bhabha Road, Mumbai 400005, India}

\date{\today}

\begin{abstract}
We present the ``trimon", a multi-mode superconducting circuit implementing three qubits with all-to-all longitudinal coupling. This always-on interaction enables simple implementation of generalized controlled-NOT gates which form a universal set. Further, two of the three qubits are protected against Purcell decay while retaining measurability. We demonstrate high-fidelity state swapping operations between two qubits and characterize the coupling of all three qubits to a neighbouring transmon qubit. Our results offer a new paradigm for multi-qubit architecture with applications in quantum error correction, quantum simulations and quantum annealing.

\end{abstract}

\maketitle

Controlling and manipulating the interactions between multiple qubits is at the heart of quantum information processing, and the superconducting circuit architecture \cite{sup-qubit-review-science} has emerged as a leading candidate. Previous demonstrations of multi-qubit devices \cite{Martinis_9xmon_nature, IBM-errordet-4qubit, riste2015detecting, multi-qubit-3d-yale,Qsim_xmon, wallraff-dig-qsim,RIPgate} have used transmon qubits \cite{transmon_theory} along with separate coupling elements to implement transverse inter-qubit coupling. Typically, this transverse coupling is weak and restricted to nearest neighbours which limits the kind of multi-qubit operations that can be performed. Recently, longitudinal inter-qubit coupling has been proposed as an alternative for building a universal multi-qubit architecture \cite{Blais-long-coup,nakamura-long-coup,divincenzo-long-coup} and for quantum annealing architectures with all-to-all coupling \cite{Transmon-annealer-zoller}. While the transmon design uses a single anharmonic oscillator mode to implement a qubit, this idea can be extended to a circuit that can support several oscillator modes to implement a multi-qubit system with strong longitudinal coupling \cite{Blais_TCQ,Auffeves_Buisson-theory}. However, previous experiments \cite{Houck-TCQ2,Buisson-Vshaped} have not demonstrated multi-qubit operations and their coherence times have not matched that of typical transmon qubits. 

\begin{figure} [t]
	\includegraphics[width=\columnwidth]{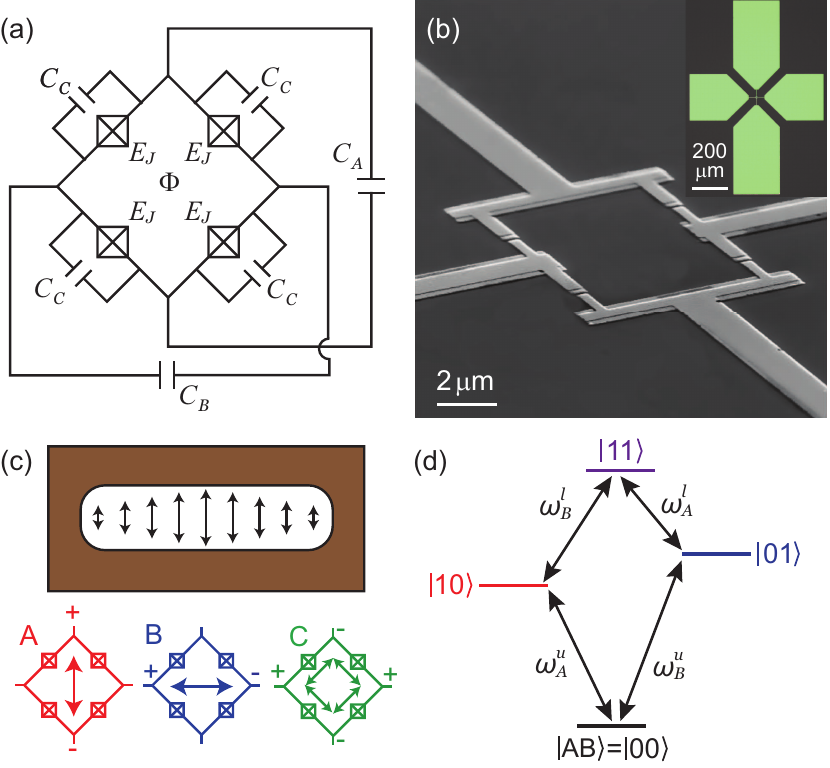}
	\caption{(colour online) (a) Circuit schematic and (b) scanning electron microscope image of the trimon device. Inset: optical image (false colour) of the full device. The two pairs of capacitor pads have different sizes to obtain different transition frequencies for  qubits A and B. (c) The device is placed at the center of a rectangular copper cavity with qubit A's dipole (red arrow) aligned to the cavity's electric field in the TE$_{101}$ mode (black arrows). Qubit B's orthogonal dipole (blue arrow) and qubit C's quadrupole (green arrow) are also indicated. (d) Energy level diagram of the coupled two-qubit subspace of A and B with qubit C in its ground state. The $\sigma_z \sigma_z$ coupling makes the transition frequency of each qubit dependent on the state of the other. The upper ($\omega_{A,B}^u$) and lower ($\omega_{A,B}^l=\omega_{A,B}^u-2J_{AB}$) band transition frequencies for each qubit are indicated.}
	\label{fig:schematic}
\end{figure}

In this Letter, we present a new quantum device, the ``Trimon", implementing a three-qubit system that arises from a single superconducting circuit. Our device (Fig.~\ref{fig:schematic}(a)) is based on the Josephson ring modulator (JRM) consisting of four nominally identical Josephson junctions in a superconducting loop to implement three orthogonal electrical modes \cite{JRM}. This three-mode structure has been previously exploited to couple different harmonic oscillators for parametric amplification \cite{JPC_expt}, while more recently, it has been proposed as a coupling element between two qubits \cite{Transmon-annealer-zoller}. Here, we capacitively shunt the JRM by connecting superconducting pads to each node (Fig.~\ref{fig:schematic}(b)) to create three coupled anharmonic oscillator modes: two dipolar and one quadrupolar (Fig.~\ref{fig:schematic}(c)). Each mode has properties similar to 3D-transmon qubits \cite{paik3DT} with the resonant frequency and anharmonicity controllable by design. The longitudinal inter-qubit coupling \cite{Transmon-annealer-zoller} of the cross-Kerr type originates due to the sharing of the four junctions amongst all three modes. One of the two dipolar modes couples directly to the host 3D electromagnetic cavity (Fig.~\ref{fig:schematic}(c)); we call this the ``A" qubit. The other dipolar mode (qubit B) and the quadrupolar mode (qubit C) ideally stay uncoupled from the cavity and hence protected from Purcell decay \cite{Blais_TCQ}. However, this protection does not preclude cavity-based measurement of qubits B and C; the inter-qubit longitudinal coupling results in dispersive shifts similar to that of qubit A (Section I, Ref.~\onlinecite{supp}).

The Hamiltonian of our circuit (Section I, Ref.~\onlinecite{supp}) when operated at zero flux in the loop is given by 
\begin{multline}
	\dfrac{1}{\hbar} H_{\rm{system}} =-\dfrac{1}{2} \left[ \sum_{i=A,B,C} (\omega_i -2\beta_i) \sigma_z^i + \sum_{i\neq j}J_{ij}\sigma_z^i \sigma_z^j \right] \\
	+ \left(\omega_{\rm{cav}} -  \sum_{i=A,B,C}\chi_i \sigma_z^i \right)  a^\dagger a, 
\end{multline}
where $\omega_{i=A,B,C}$ are uncoupled qubit transition frequencies, $\beta_i = J_i+J_{ij}+J_{ki}, \ i\neq j\neq k,$ are the shifts due to self-Kerr ($J_i$) and cross-Kerr ($J_{ij}$) terms, $\chi_{i=A,B,C}$ are the dispersive shifts and $\omega_{\rm{cav}}-\sum_{i}\chi_i$ is the cavity frequency with all qubits in the ground state. While the transition frequencies can be tuned with a non-zero flux, it will introduce additional terms in the Hamiltonian (Section I, Ref.~\onlinecite{supp}) and here we focus on the zero-flux case. Due to the longitudinal $\sigma^z\sigma^z$ coupling, each qubit now has four possible values of transition frequency that depend on the state of the other two qubits. For simplicity, the level diagram shown in Fig.~\ref{fig:schematic}(d) is restricted to the two-qubit subspace spanned by qubits A and B (with qubit C frozen in its ground state) but reveals all the important features. For a given qubit, we label the transition frequency to be in the upper $(\omega_i^u)$ or lower $(\omega_i^l=\omega_{i}^u-2J_{AB})$ band when the partner qubit is in the ground or excited state respectively. A rotation on qubit A conditioned on the state of qubit B can be realized by a microwave tone at frequency $\omega_{A}^l$ (B in the excited state) or $\omega_{A}^u$ (B in the ground state). A single pulse at $\omega_{A}^l$ with the appropriate amplitude and length then implements a conventional CNOT gate \cite{fluxqubit-CNOT,Mooij-selective_darkening1} up to $-90^\circ$ phase which we call $-i$CNOT$_{BA}$. This extra phase can be accounted for in this architecture by simply shifting the rotation axis of all subsequent pulses on qubit B (Section II, Ref.~\onlinecite{supp}). 

While the always-on $\sigma_z\sigma_z$ coupling leads to simple two-qubit gates, single qubit gates become less trivial. In order to perform a single qubit rotation on B independent of the state of A, one now needs to apply pulses at both $\omega_B^l$ and $\omega_B^u$. This is similar to an NMR technique \cite{NMR_technique_Chuang}, where a single broadband pulse covering both frequencies is used. We use a multi-frequency pulse instead, due to the large ($J_{AB}/\pi=201.2$ MHz) coupling in our system. This also automatically accounts for the phase evolution in the qubit states due to the always-on $\sigma_z\sigma_z$ coupling \cite{NMR_technique_Chuang}. In general, for such an $N$-qubit system, pulses at $2^{N-1}$ different frequencies will be required to perform a single qubit gate, $2^{N-2}$ frequencies for two qubit gates, and so on. However, since the multi-frequency pulse can be generated using a single microwave source and simple modulation techniques (Section II, Ref.~\onlinecite{supp}), we envisage our device as a three-qubit building block for constructing larger quantum circuits. 
\begin{table}[b]
		\caption{Parameters and coherence properties of the trimon. The transition frequency ($\omega^u$) of each qubit with the other two qubits in their ground state is listed along with the anharmonicity ($\alpha$), relaxation time (T$_1$), Hahn echo time (T$_2^E$), dispersive shift ($\chi$) and  inter-qubit coupling strength ($J_{i,j}$). \# For qubit C, the Ramsey fringe decay time (T$_2^R$) is indicated as we were unable to get a clear Hahn echo signal (Section III, Ref.~\onlinecite{supp}).}
		
		\begin{tabular}{c c c c c c c} 
			\hline
			\hline
			Qubit & $\omega^u/2\pi$ & $\alpha/2\pi$  & T$_1$ & T$_2^E$ & $\chi/2\pi$  & $J_{i,j}/\pi$ \\
			&(GHz)&(MHz)&($\mu$s)&($\mu$s)&(MHz)&(MHz)\\
			\hline
			A & 5.5585 & 111.0 & 20.6 & 39.7 & -0.332 & $J_{A,B}/\pi=201.2$ \\ 
			
			B & 6.1470 & 116.0 & 51.4 & 64.8 & -0.376 & $J_{B,C}/\pi=253.0$ \\
			
			C & 7.0180 & 138.6 & 26.2 & 32.3$^\#$ & -0.386 & $J_{C,A}/\pi=232.0$ \\ [0ex]
			\hline
			\hline
		\end{tabular}
		
		\label{table:coherence}
\end{table}

The trimon devices were fabricated on a high resistivity intrinsic silicon chip using standard electron beam lithography and double angle evaporation of aluminium. The device, placed inside a two port copper cavity with asymmetric coupling was put inside light-tight radiation and cryoperm shields and cooled to 30~mK in a cryogen-free dilution refrigerator. The first stage amplification of the output signal was done by a near quantum-limited, lumped-element Josephson Parametric Amplifier (LJPA) \cite{Hatridge-JPA}. Details of the measurement setup are provided in Section II of Ref.~\onlinecite{supp}. The resonant frequency and linewidth of our measurement cavity (bare) were measured to be $\omega_{\rm{bare}}/2\pi=7.23 \text{ GHz}$ and $\kappa/2\pi=3.9 \text{ MHz}$ respectively. The  upper and lower band transition frequencies were extracted using Ramsey fringe experiments.

The results of spectroscopy and coherence measurements on all three qubits are tabulated in Table~\ref{table:coherence} indicating coherence properties comparable to typical 3D transmon qubits \cite{multi-qubit-3d-yale}. Note that the anharmonicities ($\alpha$) are about a factor of two smaller than the typical transmon values but can be increased by adjusting the design parameters. The inter-qubit coupling ($J_{i,j}$) numbers confirm the strong, pairwise longitudinal coupling.  We obtained the best relaxation time (T$_1$) for qubit B as it is decoupled from the cavity. The T$_{1,B} \sim 50~\mu s$ is consistent with our measurements on regular transmon qubits when they are detuned sufficiently from the cavity so that the relaxation time is not limited by Purcell decay. While we expected the T$_1$ for qubit C to be similar to that of B, we observed it to be smaller. This trend was seen across several devices and one possible reason could be the unavoidable spread in the Josephson energies of the four junctions. As a result, qubits B and C develop a small dipolar component along the cavity field leading to a finite qubit-cavity coupling (Section III, Ref.~\onlinecite{supp}). Since qubit C ($\omega_C^u /2\pi = 7.0180$ GHz) is much closer to the bare cavity frequency $(\omega_{\rm{bare}}/2\pi=7.23~\text{GHz})$ than qubit B ($\omega_B^u /2\pi = 6.1470$ GHz), even a small coupling to the cavity can reduce the T$_1$ due to Purcell decay. Further investigation is planned to understand this effect.
\begin{figure}[t]
	\includegraphics[width=\columnwidth]{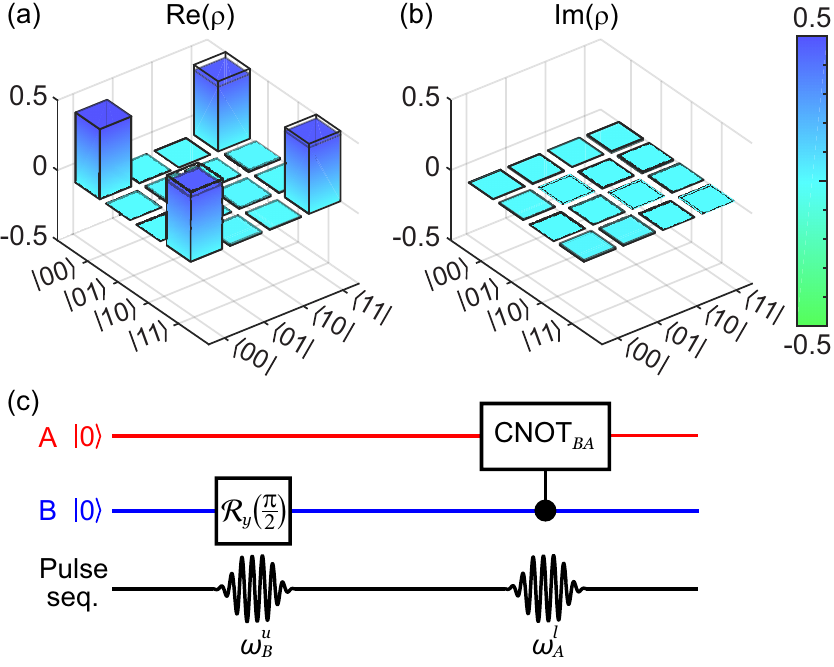}
	\caption{(colour online) (a) and (b), Real and imaginary parts of the reconstructed density matrix (MLE) for the Bell state $(|00\rangle +|11\rangle)/\sqrt{2}$ which we prepared using the single-pulse CNOT gate. Here, the filled coloured bars are experimental data while transparent bars with black boundaries denote ideal values corresponding to the intended state. (c), The quantum circuit  for preparing the Bell state and the corresponding pulse sequence with transition frequencies are indicated. The $\pi/2$-pulse length at $\omega_B^u$ was 281~ns while the $\pi$-pulse (CNOT$_{BA}$) length at $\omega_A^l$ was 241~ns.  }
	\label{fig:Bell}
\end{figure}

We first demonstrate our single-pulse CNOT gate by preparing a Bell state using qubits A and B. For all experiments, we start by performing a strong measurement and process only those data for which this measurement yields the state $|000\rangle$. A $\pi/2$-pulse at frequency $\omega_B^u$ is then applied to prepare the state $|0\rangle(|0\rangle + |1\rangle)/\sqrt{2}$. Finally, the CNOT$_{BA}$ gate is implemented by a $\pi$-pulse at $\omega_A^l$ to prepare the two-qubit Bell state $(|00\rangle +|11\rangle)/\sqrt{2}$. Note that the $\pi$-pulse at $\omega_A^l$ implements a native $-i$CNOT$_{BA}$ gate and we shift the phase of all subsequent pulses of the control qubit (B) by $90^\circ$ to construct the conventional CNOT gate. The real and imaginary parts of the reconstructed density matrix along with the pulse sequence are shown in Fig.~\ref{fig:Bell}. The fidelity of the Bell state was found to be $0.974\pm0.003$ using two-qubit tomography and maximum likelihood estimation (MLE) and does not account for finite measurement fidelity (Section V, Ref.~\onlinecite{supp}). The uncertainty quoted is the standard deviation in the fidelity in successive data sets while the uncertainty obtained from bootstrapping is about an order of magnitude lower. We prepared various other Bell states and obtained similar fidelities. 
\begin{figure}[t]
	\centering
	\includegraphics[width=\columnwidth]{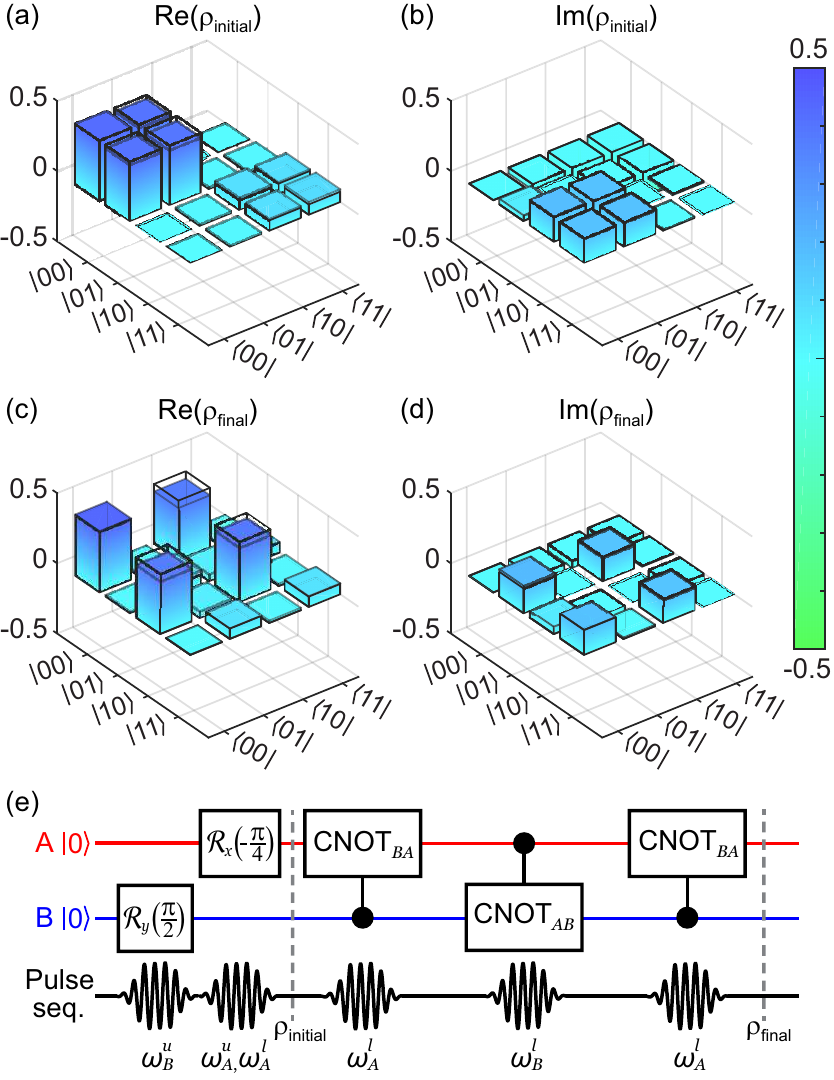}
	\caption{(colour online) Real and imaginary parts of reconstructed density matrices (a),(b) before SWAP operation; (c),(d) after SWAP operation. The filled coloured bars are the experimental data while transparent bars with black boundaries denote ideal values corresponding to the intended state. (e), The quantum circuit and the corresponding pulse sequence with transition frequencies are indicated. The $\pi/2$-pulse length at $\omega_B^u$ was 281~ns, the $\pi/4$-pulse length at $\omega_A^u$ and at $\omega_A^l$ was 108~ns, while the $\pi$-pulse lengths at $\omega_A^l$ and $\omega_B^l$ were 241~ns and 497~ns respectively. }
	\label{fig:swap}
\end{figure}

\begin{figure*}[t]
	\centering
	\includegraphics[width=1.0\textwidth]{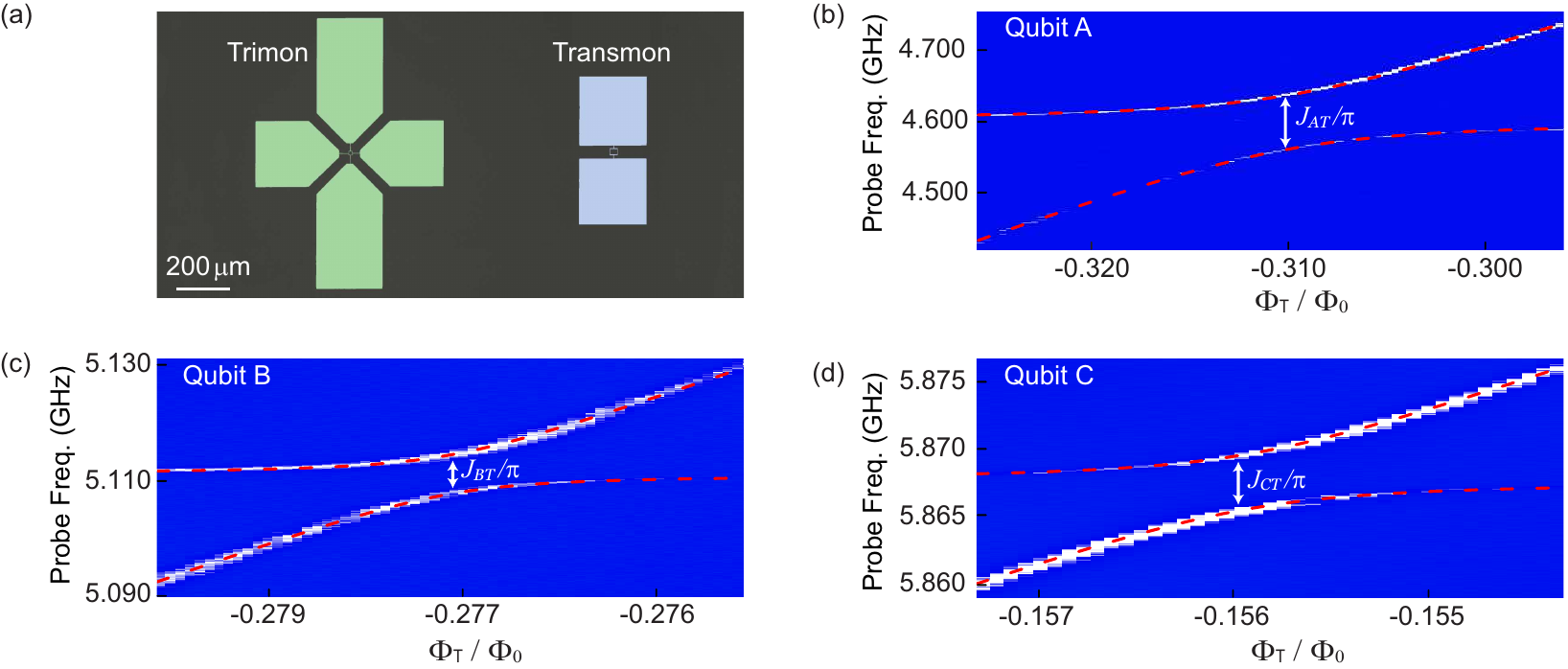} 
	\caption{(colour online) (a), False-coloured optical image of the device showing a trimon and a transmon with a spacing of about 1 mm. The dipole of the transmon is aligned with that of qubit A of the trimon device.
	(b),(c),(d) Avoided crossing of the transmon qubit's transition with qubits A, B and C. $\Phi_{\rm{T}}$ is the flux in the transmon's SQUID loop and $\Phi_0$ is the magnetic flux quantum. The dashed red lines are a fit to the location of the transition frequencies using the universal avoided crossing formula. We extract the couplings $J_{AT}/\pi=$~77.6 MHz, $J_{BT}/\pi=$~6.8 MHz and $J_{CT}/\pi=$~3.8 MHz for the three qubits respectively. 
	}
	\label{fig:Avoided}
\end{figure*}

Taking the idea further, we demonstrate an optimal SWAP gate \cite{cond-gates,optimal-SWAP}, for the first time in circuit-QED architecture, which swaps the quantum states of two qubits. Our SWAP gate is realized by three CNOT pulses on qubits A and B: SWAP$_{AB}=\text{CNOT}_{BA} \text{CNOT}_{AB} \text{CNOT}_{BA}$. We prepared the initial state $|i\rangle=(\cos(\pi/8)|0\rangle + i\sin(\pi/8)|1\rangle)_A \bigotimes ((|0\rangle+|1\rangle)/\sqrt{2})_B$ with fidelity $0.983\pm0.005$ and then using our SWAP gate generated the final state $|f\rangle=((|0\rangle+|1\rangle)/\sqrt{2})_A \bigotimes (\cos(\pi/8)|0\rangle + i\sin(\pi/8)|1\rangle)_B$ with fidelity $0.971\pm0.005$. The reconstructed density matrices of the initial and final two-qubit states are shown in Fig.~\ref{fig:swap}. If qubit B is initially in the ground state, the SWAP gate can be simplified to a transfer gate which moves a quantum state from A to B and requires only two CNOT gates. We transferred the state $(|0\rangle + |1\rangle)/\sqrt{2}$  from qubit A to B and obtained a final state fidelity of $0.973\pm0.005$ (Section VI, Ref.~\onlinecite{supp}). Further improvements in fidelities are possible by optimizing the qubit-cavity coupling and reducing pulse lengths. Given the qubit anharmonicities (Table \ref{table:coherence}) we can easily reduce the pulse lengths (see caption, Fig.~\ref{fig:Bell} and \ref{fig:swap}) by a factor of 10 without any risk of leakage out of the computational subspace. The pulse lengths in this experiment were restricted due to limited microwave power available in our setup and the relatively weak coupling of qubits to the cavity (Section III, Ref.~\onlinecite{supp}). 

Since qubits A and B  have nearly orthogonal dipole moments, their coupling to a nearby qubit, say a transmon, will depend strongly on whether the transmon's dipole is aligned \cite{kirchmair-array} to qubit A or B. To characterize the coupling between the trimon and a transmon qubit, we fabricated both the devices with their centres 1 mm apart on the same chip as shown in Fig.~\ref{fig:Avoided}(a). The SQUID loop area of the transmon was about 7 times larger than that of the trimon. This allowed us to tune the transition frequency of the transmon while keeping the transition frequencies of qubits A, B, and C relatively unchanged. Spectroscopic data in Fig.~\ref{fig:Avoided}(b)-(d) show strong coupling ($J_{AT}/\pi=$~77.6 MHz) between the transmon and qubit A since their dipoles are aligned, whereas qubit B ($J_{BT}/\pi=$~6.8 MHz) and qubit C ($J_{CT}/\pi=$~3.8 MHz) show much weaker coupling as expected. Using electromagnetic simulations, we verified that the finite coupling of qubits B and C is consistent with a $10-20 \%$ variability in the Josephson energies of the JRM arising due to fabrication uncertainties. This suggests the ability to control the trimon's coupling to a nearby qubit by moving the quantum state between qubits A and B (or C), enabling a new kind of multi-qubit architecture for quantum information. 

In conclusion, our experiments demonstrate a new multi-mode superconducting quantum circuit which implements three quantum bits with strong, pairwise longitudinal coupling. The Purcell protected qubits could potentially replace the standard transmon for many applications where a strong measurement is required without sacrificing qubit lifetime. Our architecture's native CNOT gate can be extended to three qubits to enable a single-pulse Toffoli gate \cite{toffoli-wallraff,bitflip-schoelkopf} which is crucial for quantum error correction. We demonstrate the SWAP gate which has several important applications including quantum network architectures involving flying qubits \cite{flying-SWAP}, and as an essential element of the Fredkin gate \cite{Fredkin-gate}, a universal gate suitable for reversible computing. One can also replace the standard transmon with this three-qubit unit in conventional multi-qubit architectures \cite{Martinis_9xmon_nature,IBM-errordet-4qubit, riste2015detecting} (2D or 3D) to build larger scale quantum processors with more flexibility in gate design and potentially larger gate fidelities. Finally, the trimon has the potential to be an essential building block for novel quantum computing architectures based on longitudinal coupling \cite{Blais-long-coup,nakamura-long-coup,divincenzo-long-coup}, and quantum annealing architectures with all-to-all coupling \cite{Transmon-annealer-zoller}. 

This work was supported by the Department of Atomic Energy of Government of India. R.V. acknowledges funding from the Department of Science and Technology, India via the Ramanujan Fellowship. We thank Michel Devoret, Michael Hatridge and Daniel Slichter for critical reading of our manuscript and valuable suggestions. We acknowledge Rajdeep Sensarma, Mandar Deshmukh and Nitish Mehta for useful discussions and the TIFR Nanofabrication facility.


%

\newpage
\setcounter{figure}{0}
\setcounter{table}{0}
\setcounter{equation}{0}

\global\long\def\theequation{S\arabic{equation}}
\global\long\def\thefigure{S\arabic{figure}}
\global\long\def\thetable{S\arabic{table}}

\onecolumngrid

\begin{center}
	{\bf \large Supplementary Material for ``Implementation of pairwise longitudinal coupling in a three-qubit superconducting circuit"}
\end{center}

\begin{center}	
	{Tanay Roy$^{1}$, Suman Kundu$^{1}$, Madhavi Chand$^{1}$, Sumeru Hazra$^{1}$, N. Nehra$^{1},^*$ R. \\ Cosmic$^{1},^\dagger$, A. Ranadive$^{1}$, Meghan P. Patankar$^{1}$, Kedar Damle$^{2}$, and R. Vijay$^{1\ddag}$}\\
	$^{1}$\textit{Department of Condensed Matter Physics and Materials Science, \\ Tata Institute of Fundamental Research, Homi Bhabha Road, Mumbai 400005, India and\\ $^{2}$Department of Theoretical Physics,\\ Tata Institute of Fundamental Research, Homi Bhabha Road, Mumbai 400005, India}
\end{center}
\newpage

\section{Hamiltonian Derivation}
\begin{figure}[h]
	\includegraphics[scale=1]{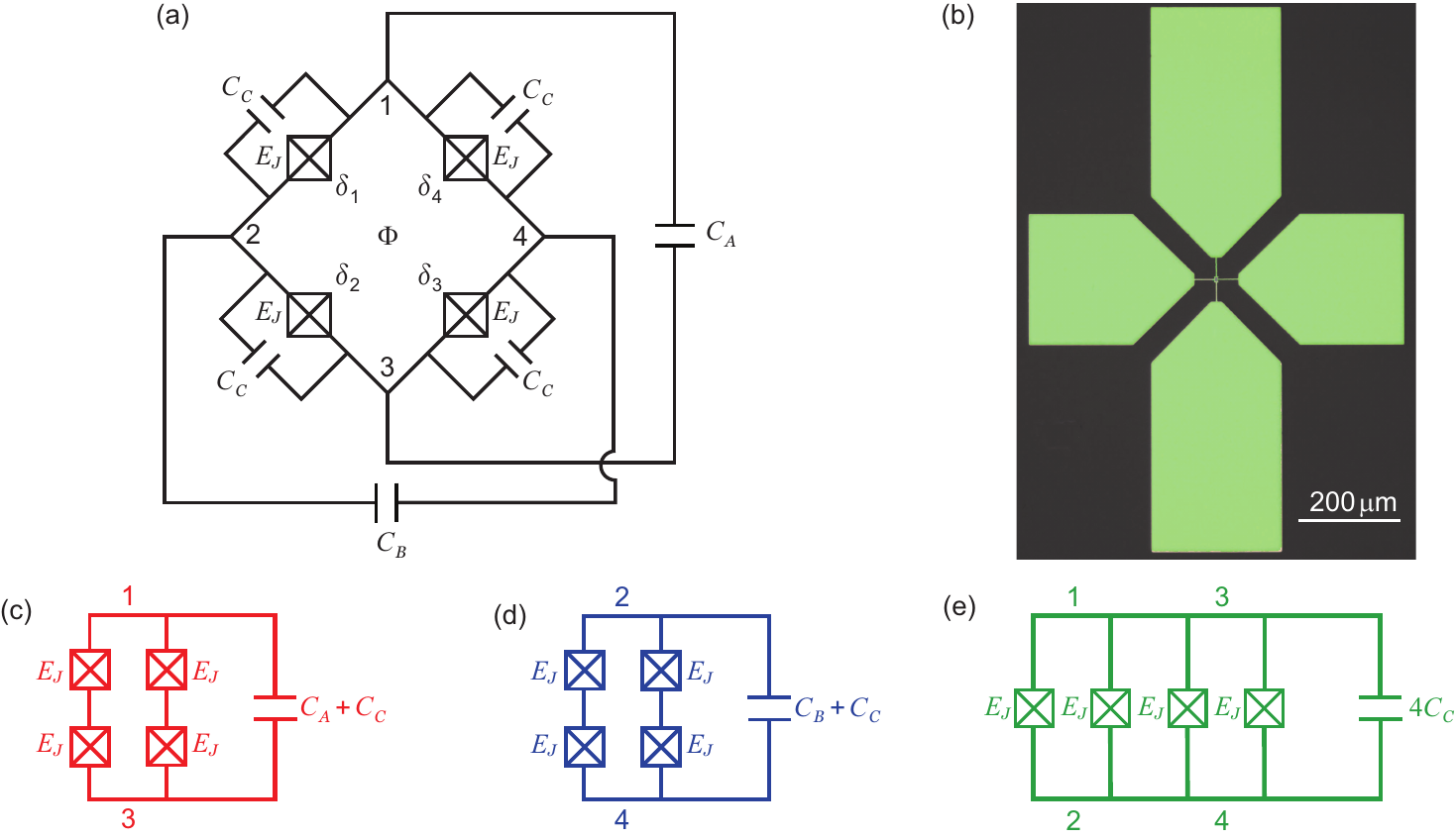}
	\caption{(a), Schematic circuit diagram of trimon device consisting of four Josephson junctions and six capacitors between four nodes. $\delta_{\mu=1,2,3,4}$ are the phase differences across the junctions with identical Josephson energies $E_J$. (b), False coloured optical image of the trimon device. (c),(d),(e), Effective circuit diagram for the A, B and C modes.}
	\label{fig:trimon_schematic}
\end{figure}

The trimon device consists of four Josephson junctions (each with Josephson enery $E_J$) in a superconducting loop with four capacitor pads connected to each node. Besides capacitances $C_C$ between adjacent node pairs (which includes the intrinsic junction capacitance), these pads give rise to capacitances $C_A$ and $C_B$ between diagonal nodes as shown in Fig.~\ref{fig:trimon_schematic}(a). This circuit provides three orthogonal oscillating modes\cite{JRM_s}: two dipolar modes and a quadrupolar mode. The dipolar mode whose electric field is aligned with that of the 3D measurement cavity (TE$_{101}$ mode) is called the A mode. The second dipolar mode B which is perpendicular to A, and the quadrupolar mode C ideally remain uncoupled from the cavity.

In order to derive the Hamiltonian for the system, we follow the approach taken by Bergeal \textit{et al.} in the JRM article\cite{JRM_s} and define node fluxes $\Phi_{\mu=1,2,3,4}$ which are related to the potentials $(V_{\mu=1,2,3,4})$ at circuit nodes  $1,2,3,4$ as
\begin{equation}
V_{\mu=1,2,3,4}=\dfrac{d\Phi_{\mu=1,2,3,4}}{dt}\equiv \dot{\Phi}_{\mu=1,2,3,4}.
\end{equation}
Note that we have used cyclic convention for node numbering which is different from that used in the JRM article\cite{JRM_s}. In terms of these node fluxes the capacitive energy in our device is
\begin{equation}
H_C = \sum_{\mu,\nu=1}^4 \dfrac{1}{2}C_{\mu \nu}\dot{\Phi}_\mu \dot{\Phi}_\nu,
\end{equation}
where the capacitance matrix $C$ is given by
\begin{equation}
C=
\begin{pmatrix}
2C_C'+C_A & -C_C' & -C_A & -C_C'\\
-C_C' & 2C_C'+C_B & -C_C' & -C_B\\
-C_A & -C_C' & 2C_C'+C_A & -C_C'\\
-C_C' & -C_B & -C_C' & 2C_C'+C_B
\end{pmatrix}.
\end{equation}
We have ignored the capacitances of each pad to the ground and $C_C'=C_C + C_J$, where $C_J$ is the intrinsic junction capacitance. The inductive energy of our circuit is
\begin{equation}
H_J = -\sum_{\mu=1,2,3,4}E_J \cos \delta_\mu,
\end{equation}
where $E_J$ is the Josephson energy of each junction (assumed identical) and $\delta_{\mu=1,2,3,4}$ are the gauge-invariant phase differences across the junctions satisfying the condition $(\delta_1+\delta_2+\delta_3+\delta_4) \mod{2\pi}=\Phi/\varphi_0$. Here $\Phi$ is the total flux threading the loop and $\varphi_0=\Phi_0/2\pi$ is the reduced flux quantum. We now write the junction phases in terms of the node fluxes
as
\begin{subequations}
	\begin{align}
	\delta_1 &= \frac{1}{\varphi_0} \left(\Phi_1-\Phi_2+\frac{\Phi}{4} \right),\\
	\delta_2 &= \frac{1}{\varphi_0} \left(\Phi_2-\Phi_3+\frac{\Phi}{4} \right),\\
	\delta_3 &= \frac{1}{\varphi_0} \left(\Phi_3-\Phi_4+\frac{\Phi}{4} \right),\\
	\delta_4 &= \frac{1}{\varphi_0} \left(\Phi_4-\Phi_1+\frac{\Phi}{4} \right).
	\end{align}
\end{subequations}

In order to transform to the mode variable ($\Phi_{i=A,B,C}$) we use the following transformations:
\begin{subequations}
	\begin{align}
	\Phi_A &=\Phi_1-\Phi_3,\\
	\Phi_B &=\Phi_4-\Phi_2,\\
	\Phi_C &=\frac{1}{2}(\Phi_1-\Phi_2+\Phi_3-\Phi_4).
	\end{align}
\end{subequations}
Note that, the C mode amplitude is defined differently \cite{JRM_s} here so that the Hamiltonian of qubit C is identical to that of a standard transmon. Then the Josephson energy $H_J$ of the circuit \cite{JRM_s} simplifies to
\begin{equation}
\label{eq:Hring}
H_{J}= -4E_J\bigg[\cos\left(\frac{\Phi_A}{2\varphi_0}\right)\cos\left(\frac{\Phi_B}{2\varphi_0}\right)\cos\left(\frac{\Phi_C}{\varphi_0}\right)\cos\left(\frac{\Phi}{4\varphi_0}\right) + \sin\left(\frac{\Phi_A}{2\varphi_0}\right)\sin\left(\frac{\Phi_B}{2\varphi_0}\right)\sin\left(\frac{\Phi_C}{\varphi_0}\right)\sin\left(\frac{\Phi}{4\varphi_0}\right)\bigg],
\end{equation}
while the capacitive energy can be expressed as
\begin{equation}
\sum_{i,=A,B,C} \dfrac{1}{2} \dfrac{e^2}{2E_{C_i}}\dot{\Phi}_i^2,
\end{equation}
where $e$ is the electronic charge and the charging energies $E_{C_{i=A,B,C}}$ are given by
\begin{equation}
E_{C_A}=\dfrac{e^2}{2(C_C'+C_A)}, \ E_{C_B}=\dfrac{e^2}{2(C_C'+C_B)}, \ E_{C_C}=\dfrac{e^2}{8C_C'}.
\end{equation}

We now express the Hamiltonian of the full system using mode charge variables $Q_{i=A,B,C}$ (which are canonically conjugate to mode flux variable $\Phi_{i=A,B,C}$) for each mode and corresponding charging energies $E_{C_{i=A,B,C}}$
\begin{equation}
\label{eq:Hcircuit}
H_{\rm{circuit}}=H_J+\sum_{i=A,B,C}E_{C_i} \dfrac{Q_i^2}{e^2},
\end{equation}
The magnitudes of $C_{i=A,B}$ are chosen to be unequal to lift the degeneracy between qubits A and B. Operating at zero applied flux $(\Phi=0)$, we expand Eq. \eqref{eq:Hcircuit} up to fourth order in the mode amplitudes to get
\begin{equation}
\begin{split}
H_{\rm{circuit}}& = -4E_J + \left( E_{C_A}q_A^2 + \dfrac{E_J}{2}\phi_A^2 -  \dfrac{E_J}{96}\phi_A^4 \right) 
+ \left( E_{C_B}q_B^2 + \dfrac{E_J}{2}\phi_B^2 -  \dfrac{E_J}{96}\phi_B^4 \right)
+ \left( E_{C_C}q_C^2 + \dfrac{4E_J}{2}\phi_C^2 -  \dfrac{E_J}{6}\phi_C^4 \right)\\
& \quad -\dfrac{E_J}{16}\left( \phi_A^2\phi_B^2 + 4\phi_B^2\phi_C^2 + 4\phi_C^2\phi_A^2  \right),
\end{split}
\label{eq:Htrunc}
\end{equation}
where $q_i=Q_i/e$ and $\phi_i=\Phi_i/\varphi_0$. Here each qubit is expressed as a weakly anharmonic oscillator (transmon). While qubit C looks like a regular transmon, the qubits A and B show reduced non-linearity. This is because in qubit C (quadrupole), all the junctions are effectively in parallel, while qubits A and B have two junctions in series (Fig.~\ref{fig:trimon_schematic}(c)-(e)). As we will see later, this dilutes the anharmonicity of A and B by a factor of 4. We now quantize the circuit by introducing the bosonic raising and lowering operators which are related to the flux and charge operators as
\begin{subequations}
	\begin{align}
	\Phi_i &= \ \sqrt{\dfrac{\hbar Z_i}{2}}(a_i^\dagger + a_i),\\
	Q_i &= i \sqrt{\dfrac{\hbar}{2Z_i}}(a_i^\dagger - a_i).
	\end{align}
\end{subequations}
We define uncoupled mode frequencies and mode impedances as,
\begin{subequations}
	\begin{align}
	\omega_A &= \dfrac{\sqrt{8E_JE_{C_A}}}{\hbar}, \ \ Z_A=\dfrac{\hbar}{e^2}\sqrt{\dfrac{E_{C_A}}{2E_J}},\\
	\omega_B &= \dfrac{\sqrt{8E_JE_{C_B}}}{\hbar}, \ \ 
	Z_B=\dfrac{\hbar}{e^2}\sqrt{\dfrac{E_{C_B}}{2E_J}},\\
	\omega_C &= \dfrac{\sqrt{32E_JE_{C_C}}}{\hbar}, \ \ 
	Z_C=\dfrac{\hbar}{e^2}\sqrt{\dfrac{E_{C_C}}{8E_J}}.
	\end{align}
\end{subequations}
Then the effective Hamiltonian under rotating wave approximation becomes,
\begin{equation}
\begin{split}
\dfrac{1}{\hbar} H_{\rm{eff}} = \sum_{i=A,B,C}\left[ (\omega_i -\beta_i) a_i^\dagger a_i -J_i a_i^\dagger a_i a_i^\dagger a_i \right]
- \sum_{i\neq j} 2J_{ij}a_i^\dagger a_i a_j^\dagger a_j,
\end{split}
\end{equation}
with
\begin{subequations}
	\label{eq:Jij}
	\begin{align}
	&\beta_i = J_i+J_{ij}+J_{ki},\ \ i\neq j\neq k, \\
	& J_A = \dfrac{E_{C_A}}{8\hbar}, \ J_B = \dfrac{E_{C_B}}{8\hbar}, \ J_C = \dfrac{E_{C_C}}{2\hbar},\\
	& J_{AB}  = \dfrac{\sqrt{E_{C_A}E_{C_B}}}{4\hbar}, \ 
	J_{BC} = \dfrac{\sqrt{E_{C_B}E_{C_C}}}{2\hbar}, \ 
	J_{CA} = \dfrac{\sqrt{E_{C_C}E_{C_A}}}{2\hbar},
	\end{align}
\end{subequations}
where $J_i$ and $J_{ij}$ are the coupling coefficients for self-Kerr ($\Phi_i^4$) and pairwise cross-Kerr ($\Phi_i^2\Phi_j^2$) terms respectively. Using second order perturbation theory one finds the energy eigenstates of the system to be
\begin{equation} \label{eq:energy_levels}
\begin{split}
\dfrac{1}{\hbar} E_{n_A,n_B,n_C} =& \sum_{i=A,B,C} \left[ (\omega_i - \beta_i) n_i - J_i n_i^2 \right] - \sum_{i\neq j} 2J_{ij}n_i n_j.
\end{split}
\end{equation}
Using Eq.~\eqref{eq:energy_levels} we compute the anharmonicities \cite{transmon_theory_s} of individual modes:
\begin{equation}
\label{eq:anharm}
\alpha_A = -\dfrac{E_{C_A}}{4\hbar}, \ \ 
\alpha_B = -\dfrac{E_{C_B}}{4\hbar}, \ \ 
\alpha_C = -\dfrac{E_{C_C}}{\hbar}.
\end{equation}
As mentioned earlier, only the anharmonicity of the qubit C is identical to that of a transmon, while those of qubits A and B are diluted by a factor of 4. However, in this device, $E_{C_C} \sim E_{C_{A,B}}/4$ and hence the anharmonicities of all three qubits are similar. 

Restricting ourselves to the four lowest energy eigenstates of Eq.~\eqref{eq:Hsys}, we write the Hamiltonian in terms of Pauli spin matrices as
\begin{equation}
\label{eq:Hspin}
\begin{split}
\dfrac{1}{\hbar} H_{\rm{spin}} = -\dfrac{1}{2} \left[ \sum_{i=A,B,C} (\omega_i-2\beta_i) \sigma_z^i + \sum_{i\neq j}J_{ij}\sigma_z^i \sigma_z^j \right],
\end{split}
\end{equation}
where
\begin{equation}
\beta_i = J_i + J_{ij} + J_{kj}, \ \ i\neq j \neq k.
\end{equation}
As a result each qubit now has 4 possible values of transition frequency depending upon the state of its partner qubits:
\begin{subequations}
	\begin{align}
	\omega_A^{B=s,C=t} &= \omega_A-2\beta_A + (-1)^s J_{AB}+ (-1)^tJ_{CA}, \ \ s,t\in\{0,1\} \\
	\omega_B^{C=s,A=t} &= \omega_B-2\beta_B + (-1)^s J_{BC}+ (-1)^tJ_{AB}, \ \ s,t\in\{0,1\} \\
	\omega_C^{A=s,B=t} &= \omega_C-2\beta_C + (-1)^s J_{CA}+ (-1)^tJ_{BC}, \ \ s,t\in\{0,1\}
	\end{align}
\end{subequations}
Since qubit C was kept in its ground state in our experiment, qubits A and B had only two transition frequencies each. We call them the upper ($\omega_{\rm{A,B}}^u$) and lower ($\omega_{\rm{A,B}}^l$) band frequencies (see Fig. 1(d) in the main text) and are given by
\begin{subequations}
	\begin{align}
	\omega_A^u = \omega_A^{B=0,C=0} &= \omega_A -2J_A - J_{AB} - J_{CA}, \\
	\omega_A^l = \omega_A^{B=1,C=0} &= \omega_A -2J_A - 3J_{AB} - J_{CA}, \\
	\omega_B^u = \omega_B^{C=0,A=0} &= \omega_B -2J_B - J_{AB} - J_{BC}, \\
	\omega_B^l = \omega_B^{C=0,A=1} &= \omega_B -2J_B - 3J_{AB} - J_{BC}.
	\end{align}
\end{subequations}

Including the interaction with the host cavity, we write the Hamiltonian of the full system as
\begin{equation}
\label{eq:Hsys}
H_{\rm{system}}=  -\dfrac{1}{2} \left[ \sum_{i=A,B,C} (\omega_i-2\beta_i) \sigma_z^i + \sum_{i\neq j}J_{ij}\sigma_z^i \sigma_z^j \right] + \hbar \left(\omega_{\rm{cav}} -  \sum_{i=A,B,C}\chi_i \sigma_z^i \right) a^\dagger a,
\end{equation}
where the dispersive shifts \cite{transmon_theory_s} for the three qubits are given by
\begin{subequations}
	\label{eq:chi}
	\begin{align}
	\chi_A &=g^2 \left(\dfrac{1}{\Delta_0}-\dfrac{1}{\Delta_1} \right),\\
	\chi_B &= \dfrac{g^2}{2} \left(\dfrac{1}{\Delta_0}-\dfrac{1}{\Delta_0+2J_{AB}} \right),\\
	\chi_C &= \dfrac{g^2}{2} \left(\dfrac{1}{\Delta_0}-\dfrac{1}{\Delta_0+2J_{CA}} \right).
	\end{align}
\end{subequations}
Here $g$ is the coupling between qubit A and cavity, $\omega_{\rm{cav}}=\omega_{\rm{bare}}-g^2/\Delta_0+\sum_i\chi_i$ with $\omega_{\rm{bare}}$ being the bare resonant frequency of the cavity, $\Delta_0=\omega_A^u-\omega_{\rm{bare}}$, and $\Delta_1=\Delta_0+\alpha_A$. Although each of the qubits has its own dispersive shift on the cavity, their origins are quite different. While qubit A has the usual dispersive shift \cite{transmon_theory_s} due to coupling to the cavity, one expects that the dispersive shifts of qubits B and C should be zero since they are completely decoupled from it. However, when qubit B (C) is excited, the transition frequency of qubit A is shifted via the $\sigma_z^A \sigma_z^B$ ($\sigma_z^C \sigma_z^A$) term. This leads to a dispersive shift in the cavity frequency since the detuning between qubit A and the cavity changes. Interestingly, for our typical device parameters, this indirect dispersive shift for B (C) qubit is similar in magnitude to the regular dispersive shift for qubit A.

It is possible to tune the transition frequencies of the three qubits by threading a non-zero flux through the JRM loop. However, the second term of Eq.~\eqref{eq:Hring} will be non-zero and introduce additional inter-qubit coupling terms. The dominant term will be the pure mixing term ($\propto \Phi_A\Phi_B\Phi_C$) which has previously been exploited for parametric amplification\cite{JRM_s,JPC_expt_s,JPC_circulator_s}. We plan to explore this regime of operation in the future where one can do gate operations by parametric pumping techniques. In practice, we have been able to tune the qubit transition frequencies down by about $200~\text{MHz}$ before the JRM experiences a jump to a different flux branch as expected and makes the device unstable. If one needs larger frequency tuning in such a device, the single Josephson junctions can be replaced with SQUIDs with a smaller loop area than the JRM. This way one can tune the effective $E_J$ while operating at integer flux quantum in the JRM loop.
\begin{center}
	\begin{table} [b]		
		\begin{tabular}{| c| c| c| c| c| c| c| c|} 
			\hline
			Technique & $J_{AB}/\pi$ & $J_{BC}/\pi$ & $J_{CA}/\pi$  & $\chi_A/2\pi$ & $\chi_B/2\pi$ & $\chi_C/2\pi$  \\ 
			\hline
			Theory (in MHz) & 227.0 & 253.6 & 248.0 &  -0.332 & -0.279 & -0.317\\ 
			\hline
			Expt. (in MHz) & 201.2 & 253.0 & 232.0 &  -0.332 & -0.376 & -0.386 \\ [0ex]
			\hline
			
		\end{tabular}
		\caption{Comparison of various device parameters between theory and experiment.}
		\label{table:comparison}
	\end{table}
\end{center}

Experimentally measured anharmonicities (see Table I in main text) are used to obtain the values of  $E_{C_{i=A,B,C}}$ using Eq.~\eqref{eq:anharm} and hence the values of couplings $J_{ij}$ using Eq.~(\ref{eq:Jij}c). Dispersive shifts of the qubits are calculated using Eqs.~\eqref{eq:chi}. The comparison between theoretical and experimentally obtained values of these parameters for sample D1 are given in Table \ref{table:comparison}. The agreement between theory and experiment for $J_{ij}$ is quite reasonable given that we have not accounted for the variability in the Josephson energies ($E_J$) of the four junctions in the JRM. These variations will introduce additional terms in the Hamiltonian and result in small changes in theoretical predictions. The disagreement for $\chi_{B,C}$ is much larger and requires further investigation. We noticed similar agreement between theory and experiment across several samples. The effect of variability in $E_J$ and the inclusion of higher order spin interaction terms will be the subject of a future manuscript.

\section{Measurement setup and pulse generation technique}

\begin{figure}[h]
	\includegraphics[width=1\textwidth]{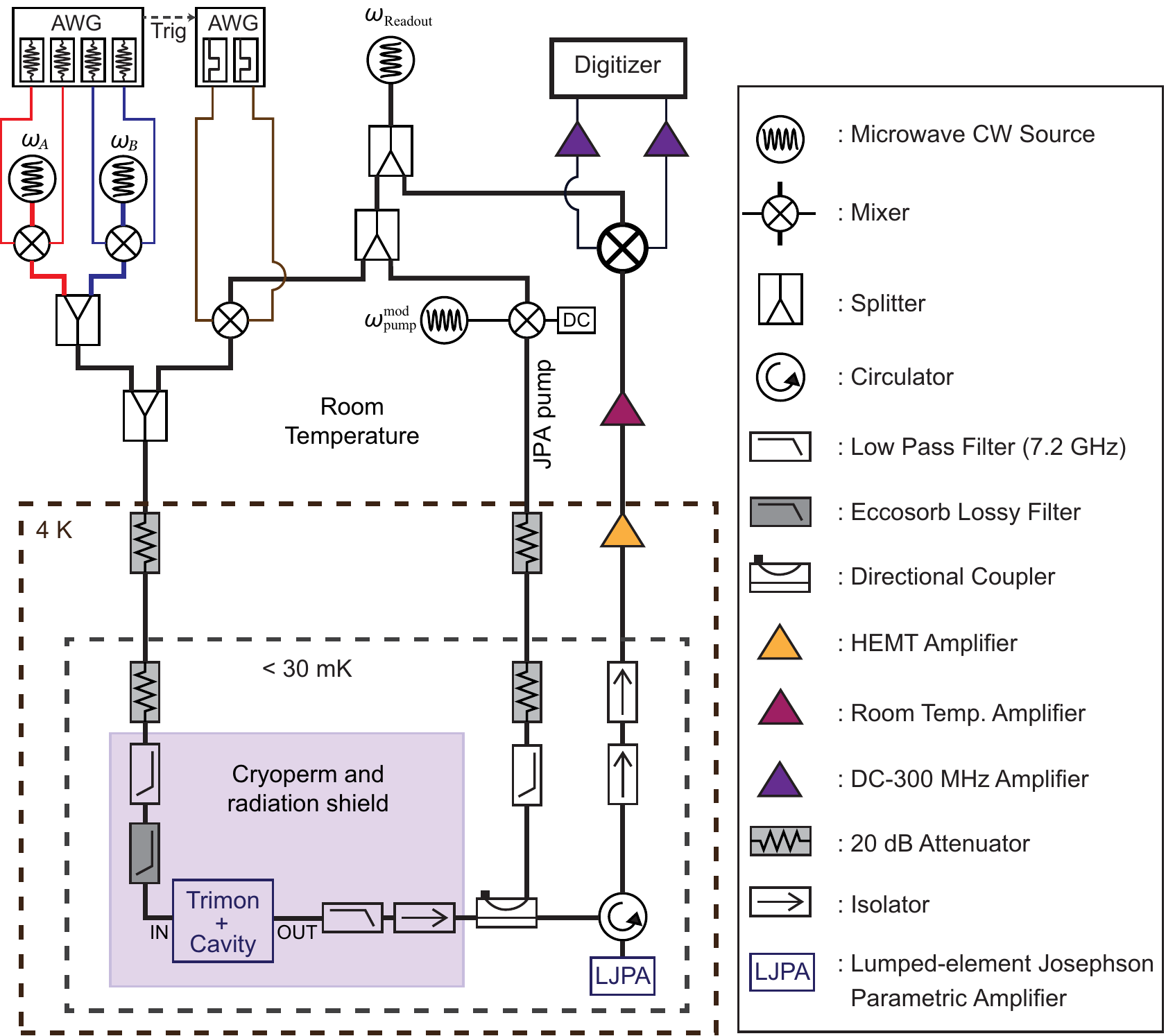}
	\caption{Room temperature signal generation and detection setup along with cryogenic wiring and filtering is shown.}
	\label{fig:trikonium_setup}
\end{figure}

Readout was performed in transmission mode where the output signal was first amplified by a nearly quantum limited, lumped-element Josephson parametric amplifier (LJPA) \cite{Hatridge-JPA_s} at base temperature followed by cryogenic (4K) and room temperature low noise amplifiers. Input lines are heavily attenuated and filtered using reflective low-pass and lossy ECCOSORB filters \cite{slichter-filter_s}. We used three microwave sources, two for the qubits (A and B respectively) and one for readout. The upper and lower band frequencies for each qubit was generated using sideband modulation technique where the local oscillators (LO) for the IQ mixers were set to the mean frequency: $(\omega_{i=A,B}^u+\omega_{i=A,B}^l)/2$. The quadrature (I and Q) modulating signals (at frequency $J_{AB}$) were generated using a 1 GS/s arbitrary waveform generator (AWG) with 300 MHz analog bandwidth. The amplitude and phase of these signals were optimized to create either a single tone ($\omega_{A,B}^u$ or $\omega_{A,B}^l$) signal for conditional rotations or a multi-tone signal ($\omega_{A,B}^u$ and $\omega_{A,B}^l$) for qubit selective rotations. The signal strengths on each band for a given qubit were adjusted to provide identical pulse lengths for a given rotation angle. We used Gaussian-edge, flat-top pulses to have a good balance between pulse bandwidth and pulse lengths. 

One particular advantage of this technique is the simple control of rotation axis by adjusting the phase of the modulating signal in software. Additionally, any rotation of the qubit about Z-axis doesn't require a separate gate and can be included by modifying the rotation axis of subsequent pulses. One important detail in this technique is to ensure that the phase difference between the upper and lower band tones for qubit A  must be identical to that of qubit B. This difference can arise due to unequal cable lengths at the RF output of IQ mixers before they are combined and sent to the device. However, this phase difference can be easily compensated by adjusting the phase of the quadrature modulating signals as well. This technique can easily be extended for a three qubit system where four tones per qubit would be generated using appropriate modulating signals.

\section{Coupling of three qubits to the cavity}

In the trimon with identical Josephson junctions, only qubit A couples to the cavity with coupling strength $g_A$. In order to estimate the spurious coupling of qubits B and C to the cavity ($g_B$ and $g_C$) due to junction asymmetries, we used the data shown in Fig.~4(b)-(d) in the main text for device D2. The relative coupling strengths of the different qubits to the transmon can be used as a rough estimate for their relative coupling strengths to the cavity. We obtain $g_B\sim g_A/10$ and $g_C\sim g_A/20$ which are small enough to ensure that the relaxation time for qubits B and C is not limited by the Purcell effect \cite{Blais-CPB_s} provided their transition frequencies are not too close to the cavity's resonance.

The relatively small coupling of the qubits B and C to the cavity makes it very difficult to couple power into those qubits and is the primary reason behind our gate pulse lengths being relatively long. It should be possible to carefully tailor the power coupling by introducing controlled asymmetry in the junctions or by redesigning port configuration on the cavity. The large power required to drive qubit C also resulted in a strong AC Stark shift \cite{acstark_s} and prevented us from obtaining a clear Hahn echo signal (see Table I in the main text).

\section{Matrix Representation of Conditional Rotations}
The two qubit states $|AB\rangle$ residing in a 4-dimensional Hilbert space are represented by the basis,
\begin{equation}
|00\rangle = \begin{pmatrix}1\\0\\0\\0\end{pmatrix}, \ \  
|01\rangle = \begin{pmatrix}0\\1\\0\\0\end{pmatrix}, \ \ 
|10\rangle = \begin{pmatrix}0\\0\\1\\0\end{pmatrix}, \ \
|11\rangle = \begin{pmatrix}0\\0\\0\\1\end{pmatrix}.
\end{equation}
The controlled rotations of qubit B in a plane making an angle $\phi$ with the $xz$-plane conditioned when A$=|1\rangle$ (lower band) and A$=|0\rangle$ (upper band) are given by,
\begin{equation}
\mathcal{R}_{Bl}(\phi, \theta)=
\begin{pmatrix}
1 & 0 & 0 & 0\\
0 & 1 & 0 & 0\\
0 & 0 & \cos(\theta/2) & -e^{-i\phi}\sin(\theta/2)\\
0 & 0 & e^{i\phi}\sin(\theta/2) & \cos(\theta/2)
\end{pmatrix}, \ \ 
\mathcal{R}_{Bu}(\phi, \theta)=
\begin{pmatrix}
\cos(\theta/2) & -e^{-i\phi}\sin(\theta/2) & 0 & 0\\
e^{i\phi}\sin(\theta/2) & \cos(\theta/2) & 0 & 0\\
0 & 0 & 1 & 0\\
0 & 0 & 0 & 1
\end{pmatrix}.
\end{equation}
Similarly controlled rotations for qubit A conditional on the state of qubit B are given by,
\begin{equation}
\mathcal{R}_{Al}(\phi, \theta)=
\begin{pmatrix}
1 & 0 & 0 & 0\\
0 & \cos(\theta/2) & 0 & -e^{-i\phi}\sin(\theta/2)\\
0 & 0 & 1 & 0\\
0 & e^{i\phi}\sin(\theta/2) & 0 & \cos(\theta/2)
\end{pmatrix}, \ \
\mathcal{R}_{Au}(\phi, \theta)=
\begin{pmatrix}
\cos(\theta/2) & 0 & -e^{-i\phi}\sin(\theta/2) & 0\\
0 & 1 & 0 & 0\\
e^{i\phi}\sin(\theta/2) & 0 & \cos(\theta/2) & 0 \\
0 & 0 & 0 & 1
\end{pmatrix}.
\end{equation}
These rotations can be used to implement generic two-qubit controlled unitary gates. The conventional CNOT gate (up to a $-90^\circ$ phase) becomes a special case of these controlled rotations, namely, $\mathcal{R}_{Bl}(-\pi/2,\pi)=\mathcal{R}_{Bl,x}(\pi)=-i\text{CNOT}_{AB}$ and $\mathcal{R}_{Al}(-\pi/2,\pi)=\mathcal{R}_{Al,x}(\pi)=-i\text{CNOT}_{BA}$ where $\phi=-\pi/2$ denotes rotation about x-axis.

\section{Two Qubit State Tomography}

For a two-qubit system the density matrix ($\rho$) can be constructed by a set of 16 linearly independent operators $\left\lbrace \mathcal O_i\right\rbrace$:
\begin{equation}
\label{eq:denmatop}
\rho=\sum_{i=0}^{15}c_i\mathcal O_i.
\end{equation}
The goal of tomography is to determine the set of coefficients $\left\lbrace c_i\right\rbrace$  from the expectation values of the observables $\left\lbrace \mathcal O_i\right\rbrace$. One such set is the Kronecker product of the Pauli matrices $\sigma_i$
\begin{equation}
\label{eq:denmatpauli}
\rho=\frac{1}{4}\sum_{i,j=x, y, z, 0}S_{ij}\sigma_i\otimes\sigma_j.
\end{equation}
The coefficients $S_{ij}$ are called the Stokes parameters. From trace normalization $\sigma_{00}$ is always zero. The problem then reduces to estimating the remaining 15 coefficients from the results of six single-qubit measurements of the type $\sigma_i\otimes\mathcal I$ or $\mathcal I\otimes\sigma_i$ and nine two-qubit measurements of the type $\sigma_i\otimes\sigma_j$ where $\left\lbrace i, j=x, y, z\right\rbrace$.

The generic form of our two qubit joint measurement operator can be written as:\cite{jointreadout_dicarlo_s}
\begin{equation}
\label{eq:jointmeas}
\mathcal O=\beta_0+\beta_1\sigma_z^A+\beta_2\sigma_z^B+\beta_{12}\sigma_z^A\otimes\sigma_z^B.
\end{equation}
Since our measurement operator involves both one and two qubit observables, the complete set of independent observables can be obtained by applying single qubit rotations prior to the measurements\cite{prerot_s}.

We have used standard Maximum Likelihood Estimation (MLE) technique \cite{MLE_s, MLE_Hradil_s} to reconstruct the density matrices. MLE searches in the parameter space of all physical density matrices and finds the density matrix $\rho$ which is most likely to have produced the observed experimental data $\mathcal D$ by constructing the “likelihood functional”. The likelihood functional is a probability distribution of obtaining the measured data given a state $\rho$, hence it is a function of the independent parameters characterizing the density matrix. For all physical states the density matrix can be written as the Cholesky decomposed form:
\begin{equation}
\label{eq:cholesky}
\rho=\frac{\mathcal T^{\dagger}\mathcal T}{Tr\left[\mathcal T^{\dagger}\mathcal T\right]},
\end{equation}
where $\mathcal T$ is an upper triangular matrix given by
\begin{equation}
\label{eq:choleskymat}
\mathcal T=\begin{bmatrix}
t_1 & t_5-it_6 & t_{11}-it_{12} & t_{15}-it_{16}\\
0 & t_2 & t_7-it_8 & t_{13}-it_{14}\\
0 & 0 & t_3 & t_9-it_{10}\\
0 & 0 & 0 & t_4 
\end{bmatrix}.
\end{equation}
The likelihood functional is defined as:
\begin{equation}
\label{eq:Likelihood}
\mathcal L\left(\mathcal D_k|\rho\left\lbrace t_i\right\rbrace \right)=\prod_{k=1}^{16}\mathcal P\left(\mathcal D_k|\rho\left\lbrace t_i\right\rbrace \right)=\prod_{k=1}^{16}\left[\left\langle\psi_k|\rho\left( t_i\right) |\psi_k\right\rangle \right]^{f_k}.
\end{equation}
Here, $\mathcal P\left(\mathcal D_k|\rho\left\lbrace t_i\right\rbrace \right)$ is the probability of having the measurement data $\mathcal D_k$ corresponding to $k^\text{th}$ measurement given the density matrix $\rho\left\lbrace t_i\right\rbrace$. The term $\langle\psi_k|\rho\left\lbrace t_i\right\rbrace|\psi_k\rangle$ denotes the probability of having the $k^\text{th}$ state and $f_k$ is the occurrence frequency of that state in an experiment. We can further simplify the expression if we assume Gaussian counting statistics and define the log-likelihood functional as:
\begin{equation}
\label{eq:gaussianest}
\mathbb L\left(\mathcal D_k|\rho\left\lbrace t_i\right\rbrace \right)=\log\mathcal L\left(\mathcal D_k|\rho\left\lbrace t_i\right\rbrace \right)=-\sum_{k=1}^{16}\frac{(\langle\psi_k|\rho\left\lbrace t_i\right\rbrace|\psi_k\rangle-f_k)^2}{2\langle\psi_k|\rho\left\lbrace t_i\right\rbrace|\psi_k\rangle},
\end{equation}
where we have set any proportionality constant to unity. Our goal is then to maximize this function with respect to the parameters $\left\lbrace t_i\right\rbrace$.

In our setup, the measurement is implemented by sending a microwave pulse at the cavity frequency. The signal transmitted through the cavity acquires a phase shift which is dependent on the joint state of the two qubits. This signal is first amplified by a near-quantum limited LJPA \cite{Hatridge-JPA_s} followed by more amplification using a cryogenic (4K) HEMT amplifier and a room temperature low noise amplifier. The amplified signal is then demodulated using homodyne technique and digitized. The digitized signal is integrated for 700 ns to create one measurement result ($V_p$) and repeating this process several thousand times allows us to create a population histogram as shown in Fig.~\ref{fig:dist}(a).

\begin{figure}[t]
	\includegraphics[scale=1.0]{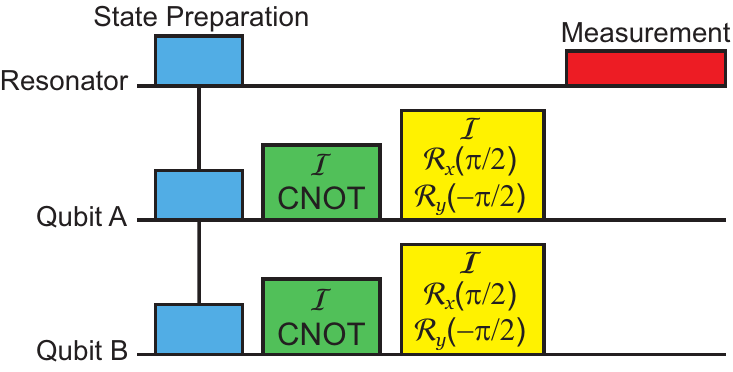}
	\caption{Pulse sequence for tomography. State preparation (blue boxes) involves strong measurement for selecting initial $|000\rangle$ state (heralding) followed by pulses to initialize the two-qubit state. The default measurement direction being the z-axis, measurements along x and y-axes for each qubit are done by performing pre-rotations by $-\pi/2$ along y-axis ($\mathcal{R}_y(-\pi/2)$) and $\pi/2$ along x-axis ($\mathcal{R}_x(\pi/2)$) respectively. Yellow boxes represent 9 possible pre-rotations and ideally should form a complete set for tomography. However, in our experiments, distributions for states $|01\rangle$ and $|10\rangle$ are largely overlapping (see Fig.~\ref{fig:dist}) and we can only distinguish states $|00\rangle$ and $|11\rangle$ with confidence. In order to differentiate $|01\rangle$ and $|10\rangle$ we repeat the same sequence with two CNOT gates prior to the pre-rotations transferring population to $|00\rangle$ and $|11\rangle$ respectively.}
	\label{fig:pulse}
\end{figure}
The occurrence frequencies $f_k$ are obtained from the population histograms with different pre-rotations as shown in Fig.~\ref{fig:pulse}. In our experimental setup the histograms corresponding to $\left|00\right\rangle$ and $\left|11\right\rangle$ are well separated from each other and also from $\left|01\right\rangle$ and $\left|10\right\rangle$, but the latter two are largely overlapping and hence indistinguishable (see Fig.~\ref{fig:dist}(a)). This is because of similar values for $\chi_A$ and $\chi_B$ (see Table \ref{table:comparison}). The measurement result is identified as the states $\left|00\right\rangle$ or $\left|11\right\rangle$ depending on whether $V_p$ is above or below some appropriately chosen thresholds $V_{\rm{th}}^+$ and $V_{\rm{th}}^-$ as shown in Fig. \ref{fig:dist}. All other values of $V_p$ between these two thresholds are discarded.  In order to obtain the population corresponding to  $\left|01\right\rangle$ and $\left|10\right\rangle$, we perform an additional measurement set by applying two CNOT gates prior to single qubit rotations. These two pulses exchange the population $\left|01\right\rangle\leftrightarrow\left|00\right\rangle$ and $\left|10\right\rangle\leftrightarrow\left|11\right\rangle$ making them distinguishable. We keep the same threshold $V_{\rm{th}}^+$ and $V_{\rm{th}}^-$ to digitize the result. The schematic of tomographic pulse sequence is shown in Fig.~\ref{fig:pulse}.
\begin{figure}[b]
	\includegraphics[scale=1.0]{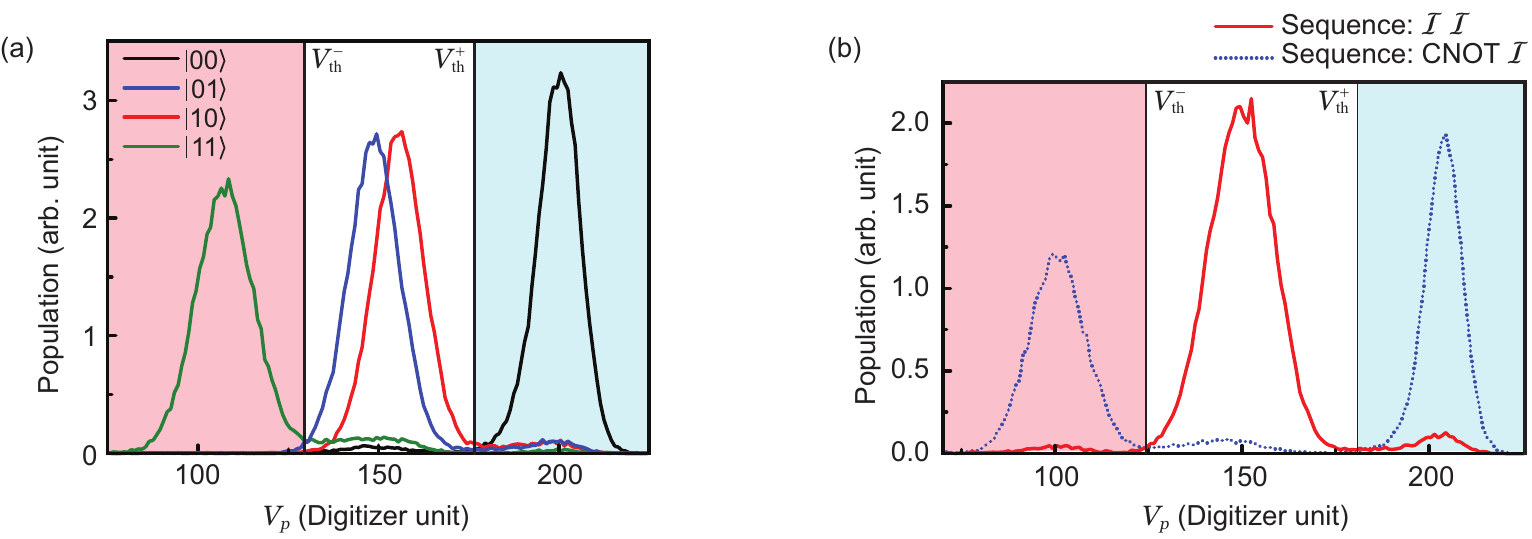}
	\caption{(a), Population histograms corresponding to individual states $\left|00\right\rangle$ (black), $\left|10\right\rangle$ (red), $\left|01\right\rangle$ (blue) and $\left|11\right\rangle$ (green) with thresholds $V_{\rm{th}}^+$ and $V_{\rm{th}}^-$. Any outcome above the threshold $V_{\rm{th}}^+$ (the light blue shaded region) will be registered as the $\left|00\right\rangle$ state. (b), Histograms of measurement results ($V_p$) for the state $(|01\rangle + |10\rangle)/\sqrt{2}$. Outcomes $V_p<V^-_\text{th}$ and $V_p>V^+_\text{th}$ are considered to be in the state $|11\rangle$ and $|00\rangle$ respectively. The red histogram shows overlapping distribution of states $|01\rangle$ and $|10\rangle$ before population transfer and the dashed blue histogram depicts the outcome after application of the CNOT gates making them distinguishable.}
	\label{fig:dist}
\end{figure}

An over-complete set of 18 pre-rotations (containing all combinations of $\left\lbrace \mathcal I,\mathcal R_x(\pi/2), \mathcal R_y(-\pi/2) \right\rbrace $ applied with and without CNOT gate) is performed to determine $\rho$. Any particular single qubit observable is measured by tracing over outcomes of the other qubit.
Since the population histograms will always have some overlap with each other, the tail of the distribution beyond the threshold will lead to wrong detection of a measurement result and eventually reduce fidelity of the reconstructed density matrix. The fidelity numbers quoted in the main text and Fig.~\ref{fig:rho_transfer} are thus limited by these finite overlaps. 

We would like to add that the density matrix is first computed by the ``Forced Purity" method \cite{forced_purity_s} (which is much faster than MLE but not very accurate) and its outcome is used to initialize the search in parameter space to obtain the density matrix $\rho_{\rm{MLE}}$. The fidelity to the target state ($\rho_{\rm{th}}$) is calculated using the formula,
\begin{equation}
\label{eq:fidelity}
\mathcal F(\rho_{\rm{th}}, \rho_{\rm{MLE}})=\rm{Tr}\left[\sqrt{\sqrt{\rho_{th}}\rho_{MLE}\sqrt{\rho_{th}}}\right].
\end{equation}

\section{Transfer gate}

\begin{figure}[t]
	\includegraphics[scale=1]{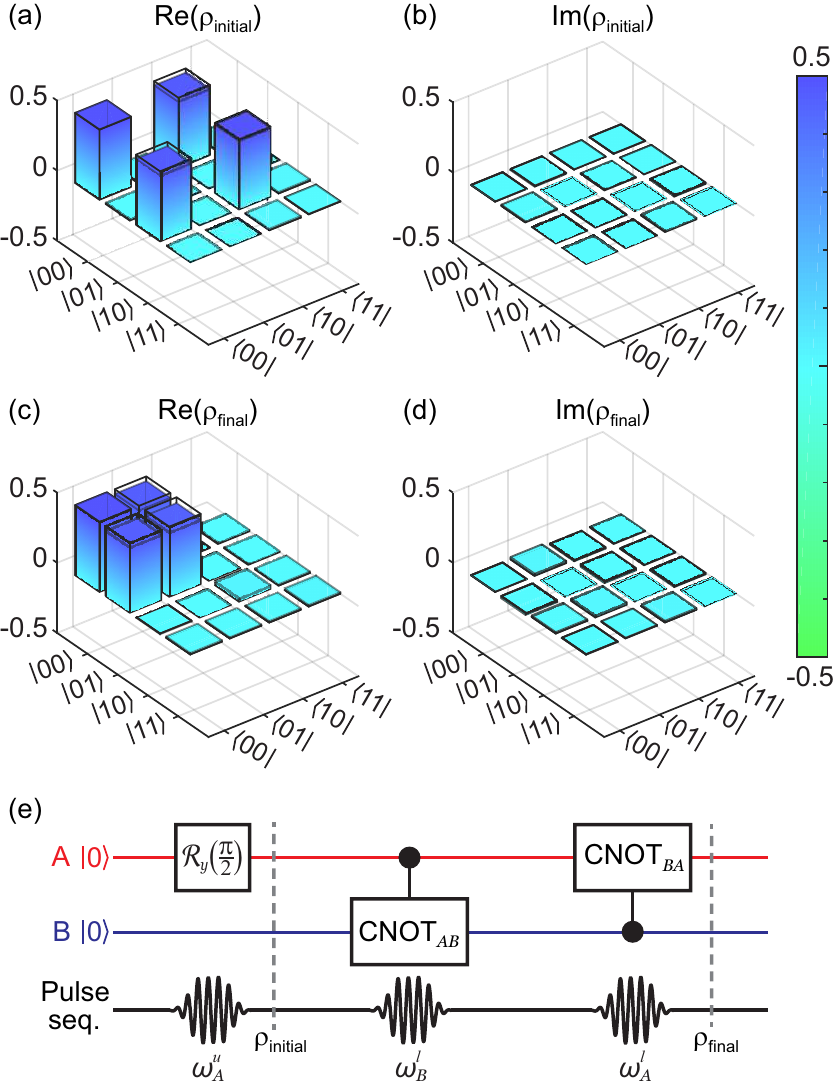}
	\caption{We initialized the two-qubit state in $((|0\rangle+|1\rangle)/\sqrt{2})_A \otimes |0\rangle_B$ with fidelity $0.984\pm0.005$ and applied transfer gate to obtain final state $ |0\rangle_A \otimes ((|0\rangle+|1\rangle)/\sqrt{2})_B$ with fidelity $0.973\pm0.005$. (a),(b) Real and imaginary components of the reconstructed density matrices before transfer; (c),(d) after transfer. Here filled coloured bars are the experimental data while transparent bars with black boundaries denote ideal values corresponding to the intended state. (e) Quantum circuit for the transfer protocol and corresponding pulse sequence. The $\pi/2$-pulse at $\omega_A^u$ was 152~ns long; pulse lengths for CNOT$_{AB}$ and CNOT$_{BA}$ were 497~ns and 241~ns respectively.}
	\label{fig:rho_transfer}
\end{figure}

Transfer gate is a special case of SWAP gate where the target qubit is initialized to its ground state and can be accomplished using only two CNOT gates on the two participant qubits. We demonstrate transfer of the state $(|0\rangle+|1\rangle)/\sqrt{2}$ from qubit A to B.  The reconstructed density matrices before and after the transfer gate along with the pulse protocol are shown in Fig.~\ref{fig:rho_transfer}. The initial state $((|0\rangle+|1\rangle)/\sqrt{2})_A \otimes |0\rangle_B$ was prepared with fidelity $0.984\pm0.005$ while the final state $ |0\rangle_A \otimes ((|0\rangle+|1\rangle)/\sqrt{2})_B$ after the transfer showed a fidelity of $0.973\pm0.005$.

\newpage

%

\end{document}